\documentclass[conference]{IEEEtran}
\IEEEoverridecommandlockouts
\usepackage{cite}
\usepackage{amsmath,amssymb,amsfonts}
\usepackage{algorithmic}
\usepackage{graphicx}
\usepackage{subfig}
\usepackage{textcomp}
\usepackage{xcolor}
\usepackage{url}
\def\BibTeX{{\rm B\kern-.05em{\sc i\kern-.025em b}\kern-.08em
    T\kern-.1667em\lower.7ex\hbox{E}\kern-.125emX}}
\begin{document}

\title{Driving innovation through project based learning: A pre-university STEAM for Social Good initiative}

\author{
\IEEEauthorblockN{Gayathri Manikutty}
\IEEEauthorblockA{\textit{AMMACHI labs} \\
\textit{Amrita Vishwa Vidyapeetham}\\
Amritapuri, India \\
0000-0003-2245-1550}
\and
\IEEEauthorblockN{Sreejith Sasidharan}
\IEEEauthorblockA{\textit{AMMACHI labs} \\
\textit{Amrita Vishwa Vidyapeetham}\\
Amritapuri, India \\
0000-0002-9049-8759}
\and
\IEEEauthorblockN{Bhavani Rao}
\IEEEauthorblockA{\textit{AMMACHI labs} \\
\textit{Amrita Vishwa Vidyapeetham}\\
Amritapuri, India \\
0000-0003-2626-1973}
}

\maketitle

\begin{abstract}
The Covid pandemic is a clarion call for increased sensitivity to the interconnected nature of social problems facing our world today. A future-oriented education on critical issues, such as those outlined in the United Nations Sustainable Development Goals (UN SDGs) and designing potential solutions for such problems is an imperative skill that must be imparted to children to help them navigate their future in today’s unpredictable world. Towards this goal, we have been conducting 3.5 month-long mentoring sessions for pre-university students in India to participate in a STEAM for Social Good innovation challenge conducted annually by the Government of India. Using digital and physical computing skills, we helped children explore creative solutions for social problems through a constructionist approach to learning, wherein they ideated and reflected upon the problems in their communities. The children learnt the Engineering Design Thinking process and worked in online groups of two or three, from concept to completion. Despite the constraints posed by the pandemic, they explored creative ways to think about design and innovation. They completed a variety of tasks by making, tinkering, engineering, assembling, and programming to grasp the intricate relationship between software and hardware. Subsequently, the children showcased their creative abilities through video storytelling to a panel of domain experts. In this paper, we present the children’s perspective of their experiences through this journey, the evaluation metrics based on IEEE design principles, and our learnings from conducting this initiative as a university-school partnership model for 84 middle and high school students. The aspirational intent of this initiative is to make the children better social problem solvers and help them perceive social problems as opportunities to enhance life for themselves and their communities. 
\end{abstract}


\section{Introduction}
Project-based learning is a constructivist, student-centred approach where students learn by completing personally meaningful projects \cite{pbl}. Constructive social problem solving is a cognitive-behavioral process to develop positive problem orientation and undertake community and societal social problem solving as a conscious, rational, effortful, and purposeful activity \cite{sps}. We adopted a pedagogical approach that purposefully situated engineering and technology knowledge and fluency within the context of social problem-solving. We created a project-based learning environment for student teams to research and gather information on societal problems, ideate solutions and create working prototypes through the engineering design process. We called our pre-university STEAM for Social Good initiative - the Amrita ATL Marathon training and mentorship program. This program was a IEEE TryEngineering STEM Program executed by IEEE Kerala Section Education Society and AMMACHI Labs, Amrita Vishwa Vidyapeetham. In this paper, we herewith present our learnings and impact results.

The shift in the Indian schooling system from an instructionist model wherein students follow tightly structured learning paths to a constructionist project-based learning model is happening at a slow pace. In the \emph{New Vision for Education: Fostering Social and Emotional Learning Through Technology} report, World Economic Forum identified sixteen critical 21st-century skills \cite{WEF}. They divided the skills into three broad categories, namely foundational literacies (literacy, numeracy, scientific literacy, Information and Communication Technology (ICT) literacy, financial literacy, cultural and civic literacy), competencies (critical thinking/problem solving, creativity, communication and collaboration), and character qualities (curiosity, initiative, persistence/grit, adaptability, leadership, social and cultural awareness). Adoption of these 21st-century skills into India’s educational standards can be accelerated only by transforming the way science and technology skills are taught and learnt, ranging from one-way content dissemination imparted by the teacher and memorization by the learner to student-centred, hands-on, meaningful learning experiences through project based learning. With this in mind, the Government of India initiated the ATL Marathon as a national level innovation challenge. This challenge is conducted annually to enhance childrens’ understanding of wicked societal problems and encourage them to engineer local solutions for global issues by delving deep into the process of designing, innovating and inventing practical solutions. Our training program focused on preparing student teams from three schools to participate in this challenge. 

Atal Tinkering Labs or ATLs are makerspaces that have been setup by the Government of India in 10,000 schools across India to create and promote a culture of innovation and entrepreneurship \cite{atl}. Our prior research with rural middle school children on integrative STEAM education workshops conducted in ATL Makerspaces has shown that such workshops improved children’s sense of agency and interest in maker projects, robotics and computational thinking \cite{akshay_steam, t4e, fie2021}. Since 2016, we have conducted several workshops in these makerspaces using innovative STEAM-based curriculum. We have observed that by incorporating arts with STEM, creativity and curiosity is fostered in children \cite{daugherty, mukil}. It allows for more human-centered innovations and encourages children to think deeply about technological development that is responsive to the needs, desires, and challenges of users \cite{akshay_usercentered, chen}. While doing projects in the makerspaces, children learn 21st-century skills of creativity, communication, collaboration and critical thinking by actively working in groups wherein they divide tasks, manage time, work together and debate upon different ideas \cite{acm, fie2021}. Our prior research findings highlight the importance of hands-on integrative STEAM education especially within a primarily instructionist education system. In our present research, we attempted to integrate engineering design thinking with STEAM for social problem solving through a 3.5 month mentorship and training program.

Dym et al. defined engineering design as a thoughtful process, stating that “Engineering design is a systematic, intelligent process in which designers generate, evaluate, and specify concepts for devices, systems, or processes whose form and function achieve clients’ objectives or users’ needs while satisfying a specified set of constraints” \cite{EDT}. They further elaborated that engineering design thinking is a complex cognitive and creative process which involves several skills. This includes the ability to “tolerate ambiguity that shows up in viewing design as inquiry or as an iterative loop of divergent-convergent thinking;  maintain sight of the big picture by including systems thinking and systems design;  handle uncertainty;  make decisions;  think as part of a team in a social process; and think and communicate in the several languages of design”. Thus by helping young children work through the engineering design process, we can cultivate a growth mindset in them to pursue challenging goals, practise thinking out of the box, embrace failures and seek advice for systematically evolving their designs. The focus of the children can be shifted from competing with others to developing openness and creating a passion for learning.

Through our intervention, we attempted to compensate for the lack of adequate practicality in traditional pedagogy and complement existing pedagogical approaches with engineering design thinking skills that are currently unaddressed for this student demographics. Our instruction method was predominantly student centered wherein students extended and refined their imbibed knowledge in developing new applications. In this study, we adopted the engineering design thinking process that was proposed by IEEE TryEngineering.org. IEEE TryEngineering.org STEM Portal is an effort to support pre-university students and teachers from around the world with resources and educational materials aimed at promoting quality STEM education \cite{tryengg_edt}. Our mentorship program followed the seven step process which included

\begin{itemize}
\item[] Step 1: Identifying the problem
\item[] Step 2: Researching the problem
\item[] Step 3: Developing possible solutions
\item[] Step 4: Selecting the best possible solution
\item[] Step 5: Constructing a prototype
\item[] Step 6: Testing and evaluating the prototype solution
\item[] Step 7: Redesigning
\end{itemize}

For this research study, our research questions were: 
\begin{enumerate}
\item What kind of impact does engineering design thinking process have on the learning opportunities for middle and high school students in India when it is incorporated into a project-based social problem solving program? 
\item What kind of changes will children see in themselves in terms of adoption of advanced technologies like artificial intelligence, internet of things, game development, robotics and web development for social problem solving? 
\item What kind of impact does such a program have on student's perceived competence, autonomy and relatedness?
\end{enumerate}

We hypothesized that children would get significant learning opportunities through our egalitarian program that would be equal regardless of gender or school locale (rural versus urban). The learning experiences we designed included opportunities for learning how technology can positively impact lives and opportunities for engaging in hands-on design challenges through advanced physical and digital computing skills. Based on this, we hypothesized that children's perceived ability to integrate advanced technology into their projects would be enhanced. We present evidence of the impact of the program on children's perceived competence, autonomy and relatedness by using the data collected and analyzed from 54 participants. We conclude this paper by sharing our learnings and provide recommendations for replicating this program across India and the world over.

\section{Theoretical Framework}

The self systems theory and self determination theory postulates that humans have three fundamental psychological needs: competence, relatedness and autonomy \cite{sdt, sst}. When children have experiences with high self-system variables, their engagement and motivation are higher \cite{TSR}. We shall now elaborate how we designed the program to boost these three needs.

\subsection{Competence} As stated earlier, the goal of our interdisciplinary pedagogy was to teach digital and physical computing concepts and tools while also providing real-world learning experiences. The 2011 report from the National Center on Time \& Learning (NCTL), \emph{Strengthening Science Education: The Power of More Time to Deepen Inquiry and Engagement} stated that when children spent more time in a school day in pursuit of STEM activities that were more hands-on and involved scientific discourses, it strengthened their competencies in STEM disciplines in a deeper way \cite{nctl}. Furthermore, Bybee stated that the development of innovative discoveries and applications requires broader and more integrative STEM educational experience wherein students have opportunities to work fluently and harmoniously across discipliness \cite{bybee}. He also added that an integrated STEM education should focus on issues related to the “global commons” such as energy efficiency, environmental quality, health maintenance and disease prevention. These issues are an integral part of social problem solving for sustainability and the United Nations Sustainable Development Goals (UN SDGs). Thus we can increase children's perceived competence by designing a cross-disciplinary pedagogical approach to teach digital and physical computing concepts while providing hands-on, real-world learning opportunities through social problem solving.

\subsection{Relatedness} The need for relatedness is a fundamental human need \cite{belong}. Interpersonal relationships can help students cope up with stress and promote positive motivational states especially during adolescent years \cite{stress}. According to the gender role socialisation perspective, girls may benefit more from close ties with teachers and mentors because closeness is congruent with greater affiliation in social relationships that are expected of girls \cite{girls_closeness}. Therefore, we made every effort to provide an encouraging atmosphere to all children during the program, wherein they could request for as many mentoring sessions as they needed for their projects. Mentors were always courteous and provided effort related praise as required.

\subsection{Autonomy} Prior research also states that when autonomy needs are met, students can develop intrinisic motivation which can, in turn influence their orientation towards different achievement goals \cite{sdt, self_worth}. According to self systems theory, autonomy is a crucial facilitator of engagement \cite{sst}. Several research studies have also shown that students' observed and self reported engagement is linked to the support for autonomy they received from teachers \cite{assor, reeve}. Therefore, we designed the program such that teachers and mentors can support these needs by showing involvement, listening more, asking more about what students wanted to do, encouraging students to ask questions, responding more to student generated questions, and taught in ways that supported student autonomy.

\subsection{Inclusivity} Another key factor in our program design was inclusivity. An inclusive STEM program is focused on preparing underrepresented youth for the successful pursuit of advanced STEM studies \cite{inclusive_stem}. Recent research studies have shown that such programs can reduce or reverse gender gaps in science attitudes and overall academic achievement \cite{reduce_gap}. We actively recruited girl students and encouraged them to form all girls' teams to participate in the program. We also actively sought students from rural regions to participate alongside their urban counterparts. Three schools participated in the program - a rural school, a semi-urban one and an urban school.

\section{Mentoring Sessions and Training Program Setup}\label{AA}
At the start of the program, students from the three schools formed their own teams of two or three students each. When students were unable to select their teammates, their teachers helped them form teams. Students could form teams of the same or mixed gender and could be an inter-school team as well. Since students studying in grades 6 to 12 were invited to participate in this program, teams could also be of mixed grade. The mentoring team consisted of eleven STEM professionals, of which six were women. Nine educators from the three schools also supported the children throughout the entire program.

The national innovation challenge organized by the Government of India called for student participation under four broad themes, namely, \emph{education}, \emph{social inclusion}, \emph{health and well-being} and \emph{energy and transportation}. The organizers provided two problem statements under each theme from which the teams could select a problem statement to work on. Alternatively, the student teams could define their own problem statement as long as the problem statement fell under one of the four broad themes. An example problem statement provided by the organizers under the theme of \emph{energy and transportation} was - \emph{``Innovate solutions that reduce the carbon footprint as well as adopt climate-resilient and low-carbon strategies to enable the transition to a truly sustainable India"}. 

\begin{figure*}[t]
\includegraphics[width=13.5cm]{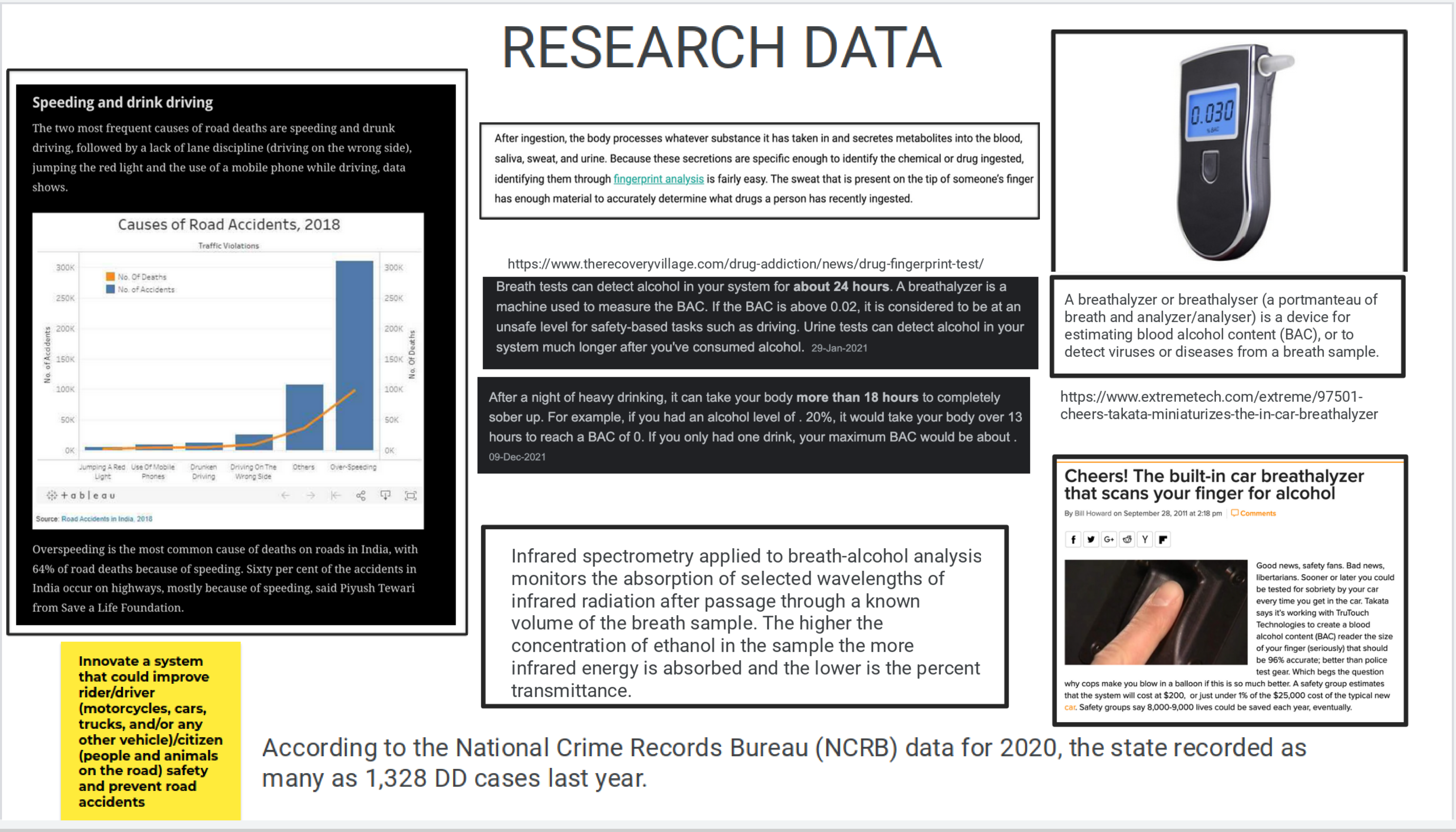}
\centering
\caption{Step 1 and Step 2: A participating team identifying and researching a problem}
\label{fig1}
\centering
\end{figure*}

\begin{figure*}[t]
\includegraphics[width=12.6cm]{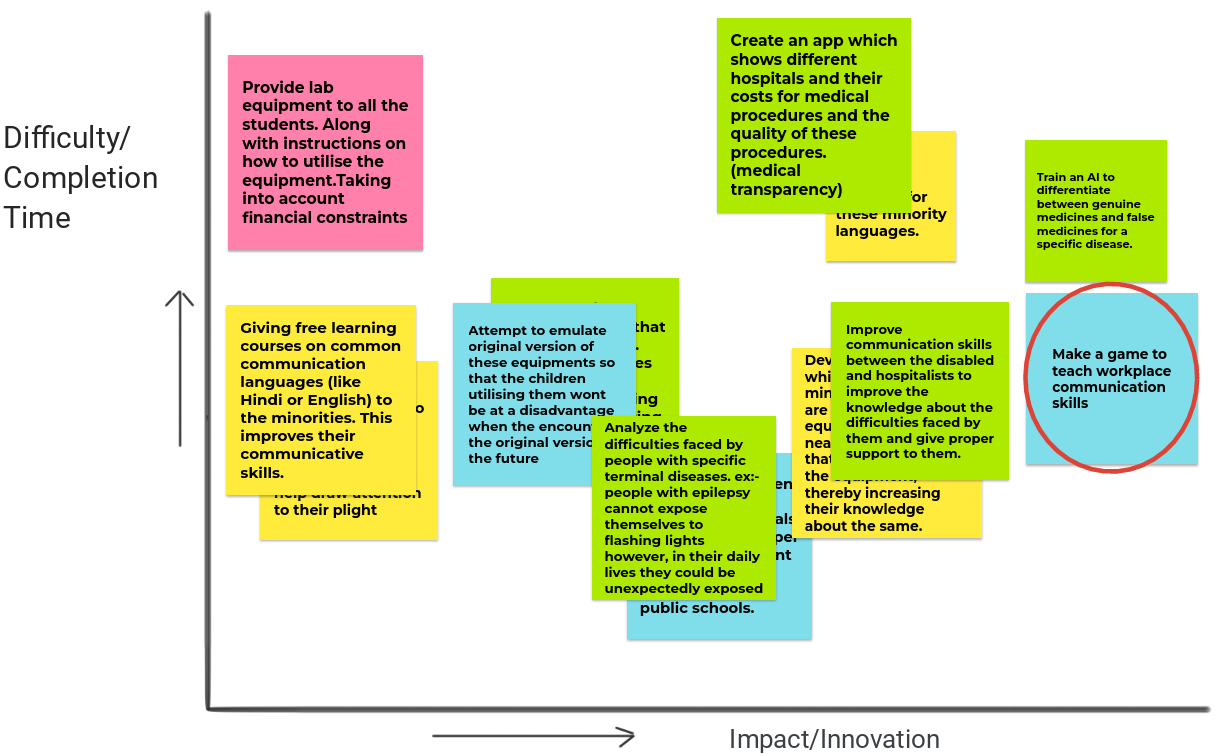}
\centering
\caption{Step 3 and Step 4: Another team's ideation exercise to select an appropriate solution}
\label{fig2}
\centering
\end{figure*}

We began the program by introducing the children to the Engineering Design Process (EDP) and showed them several examples for applying the EDP in developing and testing a prototype. The mentors then met each team individually and helped them either select one problem statement from the eight given problem statements or helped the team define their own problem statement. Then we helped the teams conduct primary and secondary research on their selected problem statement. Teams first conducted secondary research using existing data such as books, reports or research articles to understand why the problem exists and what has been done to address it in the past. Some teams then went one step further and conducted primary research to gather new data to understand who they are designing for and what kind of solution would they potentially plan on designing. After this came the solution identification step. Students practised divergent thinking to come up with several different solutions and then used convergent thinking to narrow down on the solution they wanted to finally work on. For several teams steps 1 - 4 was an iterative process. They went back and forth over the steps several times before converging on a solution. Fig.~\ref{fig1} and Fig.~\ref{fig2} shows the research and ideation process that two teams did during the program.

As each student team ideated on the problems and solutions, we asked them to document their journey using presentation slides and a 1-2 minute video capturing their chosen problem statement, the primary and secondary research undertaken and the solution they finally selected. As part of the documentation, we encouraged the teams to include an overview of their understanding of EDP and how it was applied in formulating their solution. Throughout the ideation process, mentors and school teachers provided feedback to the teams on innovation, impact, and practicality of their solutions to complete within the stipulated time and with the resources available in their school ATL makerspaces.

\begin{figure}[t]
\includegraphics[width=8cm]{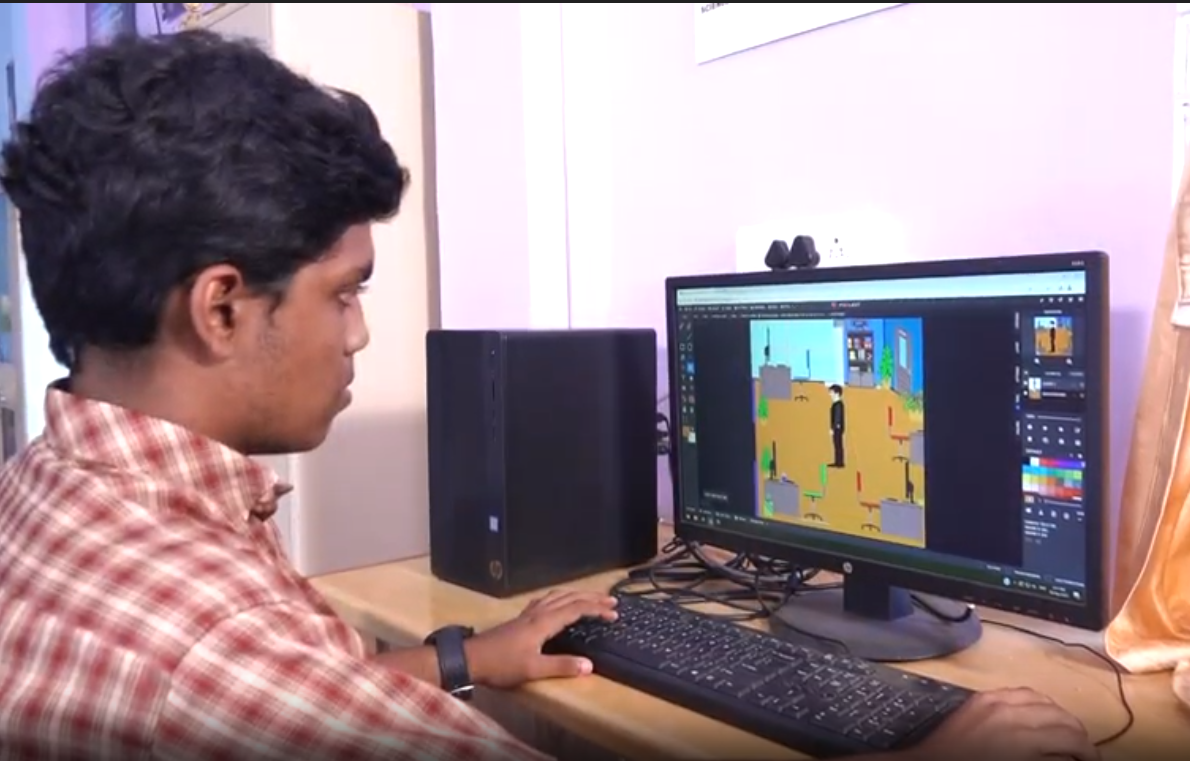}
\caption{Step 5: Student developing artwork for a game prototype.}
\label{fig3}
\centering
\end{figure}

\begin{figure}[t]
\includegraphics[width=8cm]{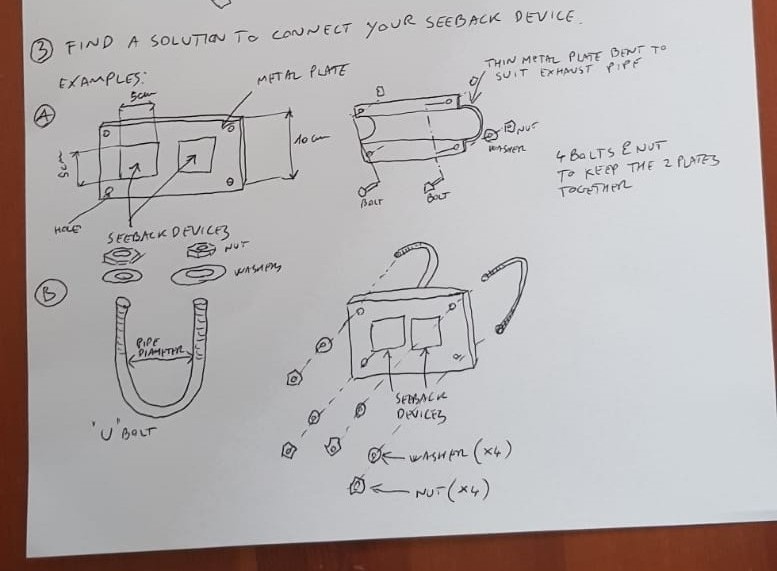}
\includegraphics[width=8cm]{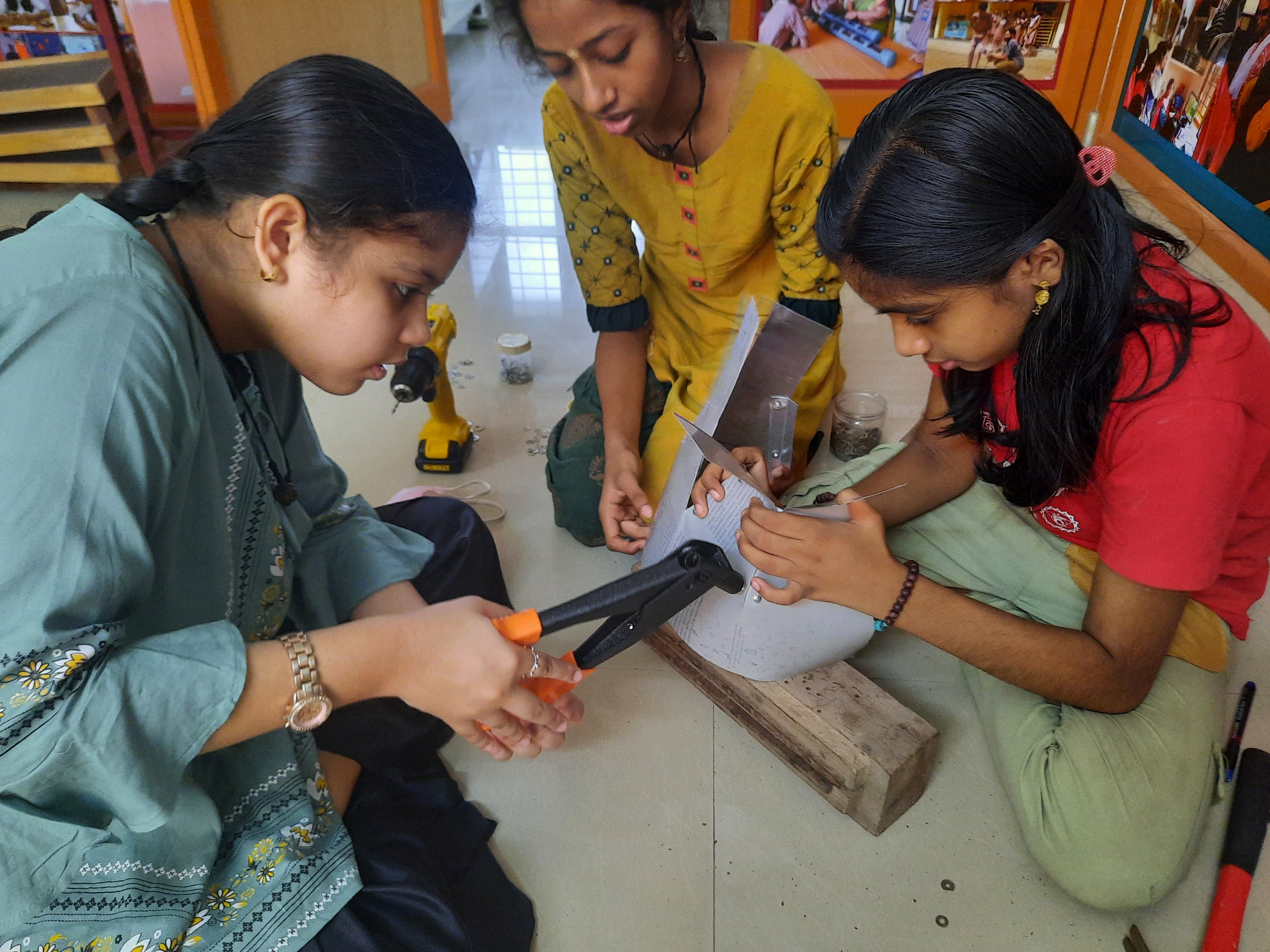}
\caption{Step 5: Students building hardware prototypes}
\label{fig4}
\centering
\end{figure}

After this step, the teams moved into the prototype development stage. Teams built either digital prototypes or physical prototypes for their solutions as shown in Fig.~\ref{fig3} and Fig.~\ref{fig4}. During this stage, mentors gathered student teams working on similar technologies and provided in depth training to the teams. In all, the mentor team conducted eight hands-on workshops covering fifteen digital and physical computing skills with varying difficulty levels. The prototype development phase extended over a two month period wherein teams had to work patiently over different kinds of challenges. Some teams had difficulty working together, with inherent disagreements and leadership issues to contend with, while some team members quit midway due to various reasons. Some teams could not get their prototype working because its difficulty level was a big leap from their level of understanding. So mentors had to take some teams back to ideation stage to select a simpler solution and start all over again. And amidst all this was also the need to handle quarantines and innumerable difficulties the Covid Pandemic brought on with it. We have documented our journey through this program at \cite{second_foundation} 

\begin{figure}[t]
\includegraphics[width=8cm]{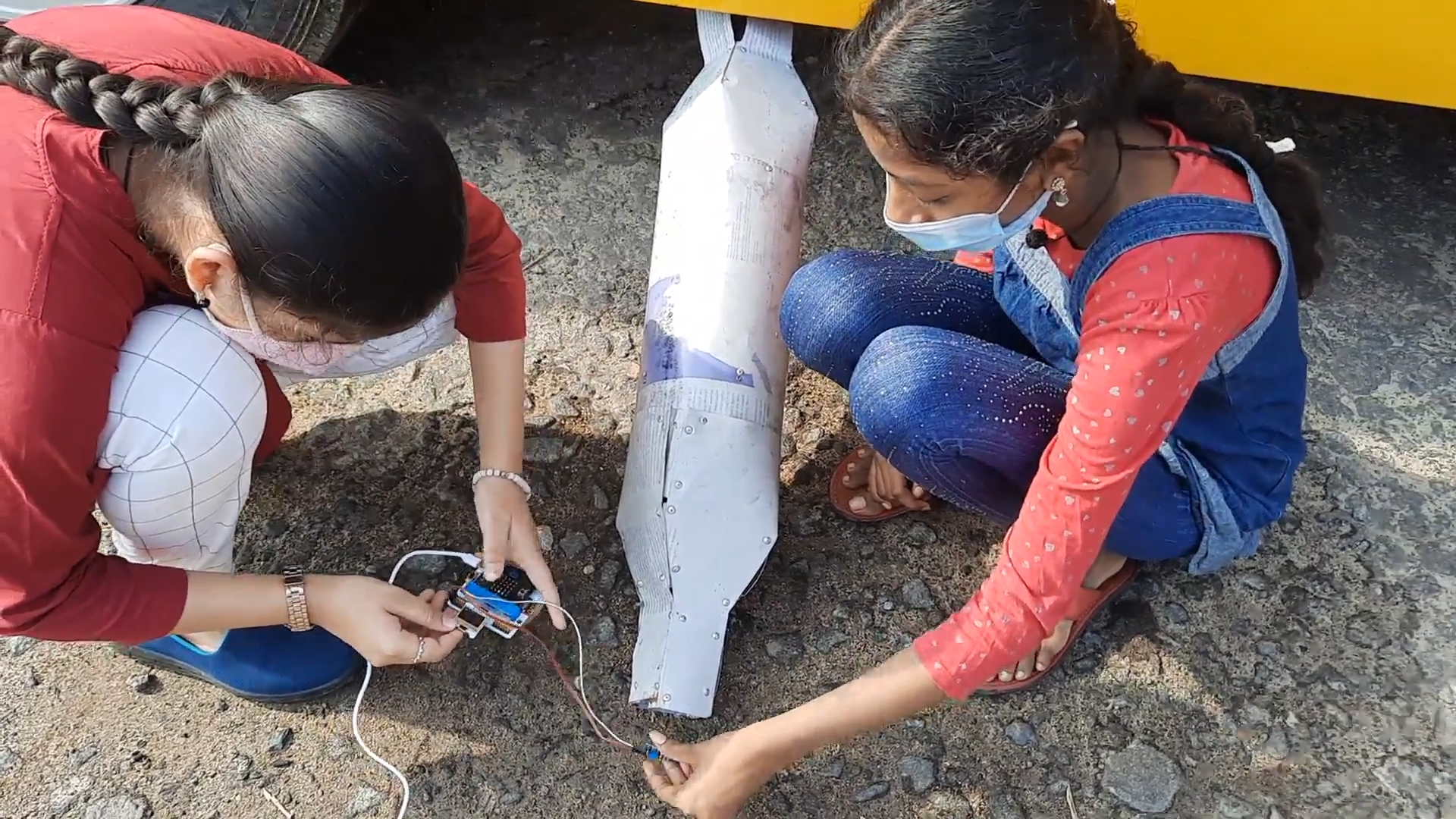}
\caption{Step 6: Students testing their prototype and taking measurements to measure efficacy of their solution.}
\centering
\end{figure}

\section{Methods}\label{sec:methods}

\subsection{Participant Profile}
We started the program with 108 children forming 40 teams from three different schools in Kerala, India. A total of 84 children (77\%)(38 girls, 46 boys) completed the program. The large attrition rate was due to the multitudinal challenges posed by the Covid pandemic, academic pressures and the poor network connectivity some children faced, especially those from rural regions. The ages of the participants ranged from 10 to 16 years. The children formed 32 teams of two or three members each with 13 all-girls' teams. Of the 84 students who completed the program, a total of 52 (22 girls) participants completed the program impact survey. Eight participants chose not to disclose their school location in the survey.

\subsection{Data Collection Instrument}

IEEE Pre-University team has formulated a set of three guiding design principles for IEEE Pre-university STEM outreach programs to have the biggest possible impact. Outlier Research and Evaluation has designed a program impact evaluation questionnaire based on the three design principles \cite{outlier}. We adapted this impact evaluation questionnaire to collect feedback from the students about our program. 

The first design principle measures the effectiveness of the program in teaching modes in which engineering can make a difference. Since our program included learning opportunities for students such as meeting stem professionals, engaging in hands-on challenges, learning technical skills and participating in solving real-world problems, we administered a questionnaire with eight items to measure all the learning opportunities the children received through the program. These included
\begin{enumerate}
\item In this program, I learnt how technology positively impacts people and communities
\item In this program, I learnt how engineers and technical professionals make an impact on society through their work 
\item In this program, I met someone who works in the STEM
\item I learnt to make something hands-on in this program	
\item In this program, I learnt new ways to solve problems 
\item In this program, I learnt about electronics hardware
\item In this program, I learnt about software 
\item In this program, I learned something about different branches of Engineering
\end{enumerate}

We measured students perceived competence based on their feedback on the learning opportunities and on one additional item, namely, 
\begin{enumerate}
\item I am more confident as a student 
\end{enumerate}

The second design principle measures how actively the program recruits youth who are in groups that are underrepresented in engineering. We measured children's impressions using the following two items.
\begin{enumerate}
\item This program showed me how people with very different backgrounds can be engineers and technical professionals 
\item This program was for people with very different backgrounds
\end{enumerate}

The third design principle measures the quality of the program design and its content. The entire program was designed for children to work in teams and encouraged collaboration on projects. Because fostering student choice and voice increases student engagement and ownership in learning, the third design principle we used was to give each student choices on what projects they wanted to do, who they will work with and how they will divide the work amongst themselves. Mentors role modelled growth mindset and provided an atmosphere of positive feedback without any judgements or comparisons. Mentors gave opportunities for students to request for guidance as many times as the student teams needed. To measure this, we included two items on students' perceived autonomy and two items on supportive relationships, that is perceived relatedness. The four items were:
\begin{enumerate}
\item In this program, I had choices in what to learn
\item In this program, I got to choose partners during learning
\item In this program, all the adults [mentors and teachers] were friendly
\item In this program, the other participants [children] were friendly
\end{enumerate}

Additionally, to measure the quality of the program content, we administered the Stages of Adoption of Technology (SA) \cite{christensen97sa} instrument to assess the improvement in stages of adoption of advanced technologies by the children. The SA instrument includes six stages which are:
\begin{itemize}
\item[] Stage 1: Awareness
\item[] Stage 2: Learning the process
\item[] Stage 3: Understanding the application of the process
\item[] Stage 4: Familiarity and confidence
\item[] Stage 5: Adaptation to other contexts
\item[] Stage 6: Creative applications to new contexts
\end{itemize}
 We asked the children to choose the stage of adoption that most appropriately described their stage of adoption of advanced technologies including artificial intelligence, internet of things, game development, robotics and web development at the beginning and at the end of the program.

\section{Results and Discussion}

In this section, we report our key findings towards the three research questions from the program impact survey administered to the participants at the end of the program.

\textbf{H1: The experience with learning opportunities will be equal regardless of gender and location.}

\begin{figure}[t]
\includegraphics[width=8cm]{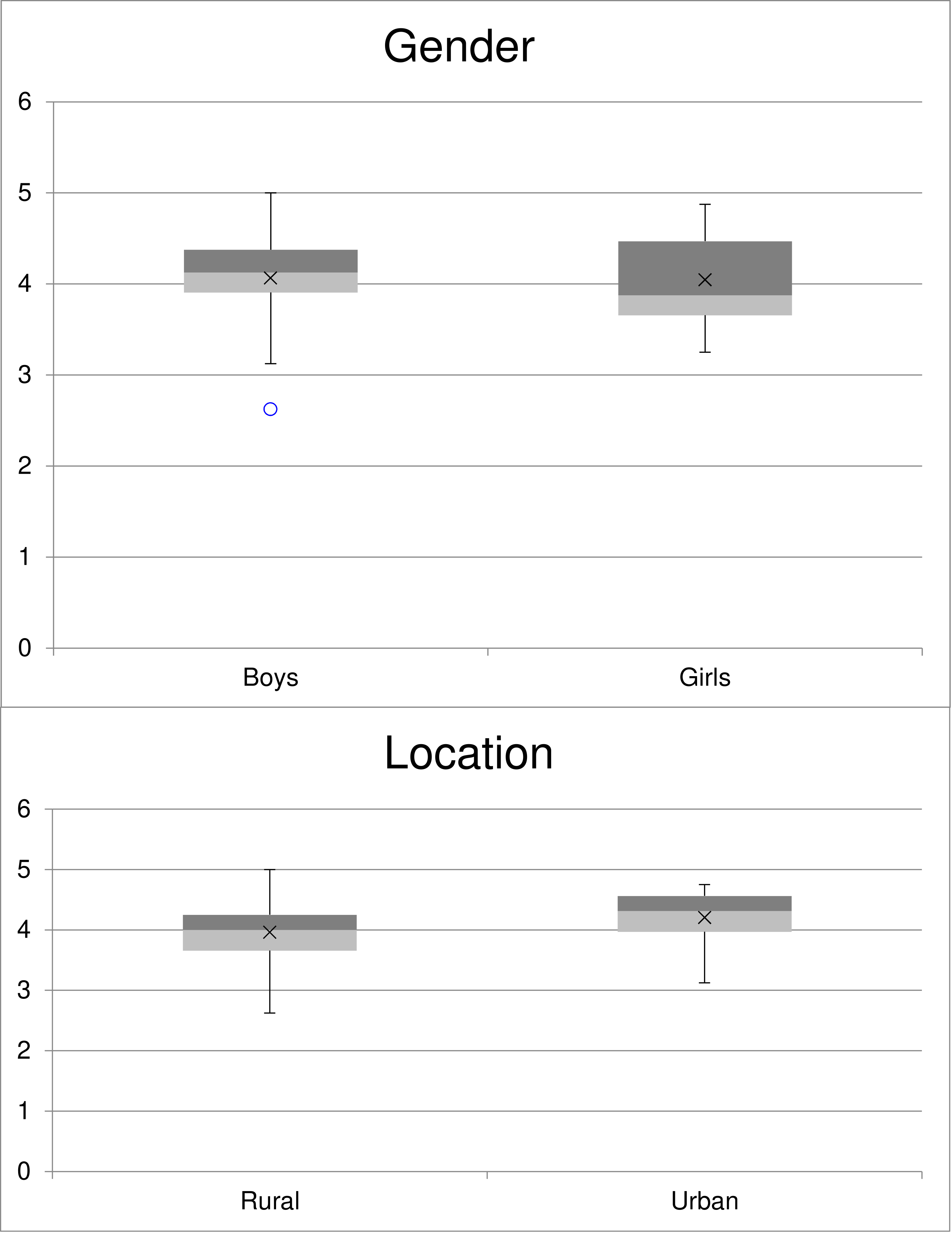}
\centering
\caption{Box plots of learning opportunities in terms of gender and school locale}
\label{fig:boxplots}
\centering
\end{figure}

All the participants believed that they had a good learning experience through this program. Fig.~\ref{fig:boxplots} shows the box plots of the means which were 4.0 or above for boys and girls and rural and urban schools. The Cronbach's alpha for the eight items was $\alpha = 0.720$. Independent samples t-test showed that there was no statistically significant difference between experiences with learning opportunities between girls (4.05 ± 0.51) and boys (4.07 ± 0.51), t(50)=0.15, p=0.88. Similarly, there was no statistically significant difference between rural (3.96 ± 0.49) and urban schools (4.21 ± 0.48), t(45)=1.65, p=0.11. These results provide evidence that the program provided equal learning opportunities for all children. We also provide some excerpts from the children's qualitative feedback on the program in support of the above.

One of the female participants from an urban school wrote \textit{``I liked the ATL marathon program [mentorship program] because it was like helping people with the use of modern technology. And I love to help people and I always wanted to know more about technology. And so this was my chance to learn much more about technology and to help people."} (Gayathry)

Another female participant from a semi urban school agreed saying \textit{``ATL marathon is a hands-on program that enable[s] students to solve problems throughout Design thinking process. From this younger age itself a child is being punctual, thinks like a mechanical engineer or a software engineer and handles project work and school studies hand in hand. It's really fun, mentors and teachers are like our friends who help us whenever in need."} (Arunima)

A male participant from a rural school wrote \textit{``I like this program very much, I learnt about new technology and how technology can help society. The most important thing about ATL marathon was the mentors. I am very much blessed to have my mentor. Without his help, we were unable to complete the project. His patience, care, continuous guidance and support made this project a successful one."} (Advaith)

\textbf{H2: Program would significantly impact children's perceived ability to integrate advanced technology into their projects and will impact equally regardless of gender and location.}

\begin{figure}[t]
\includegraphics[width=8cm]{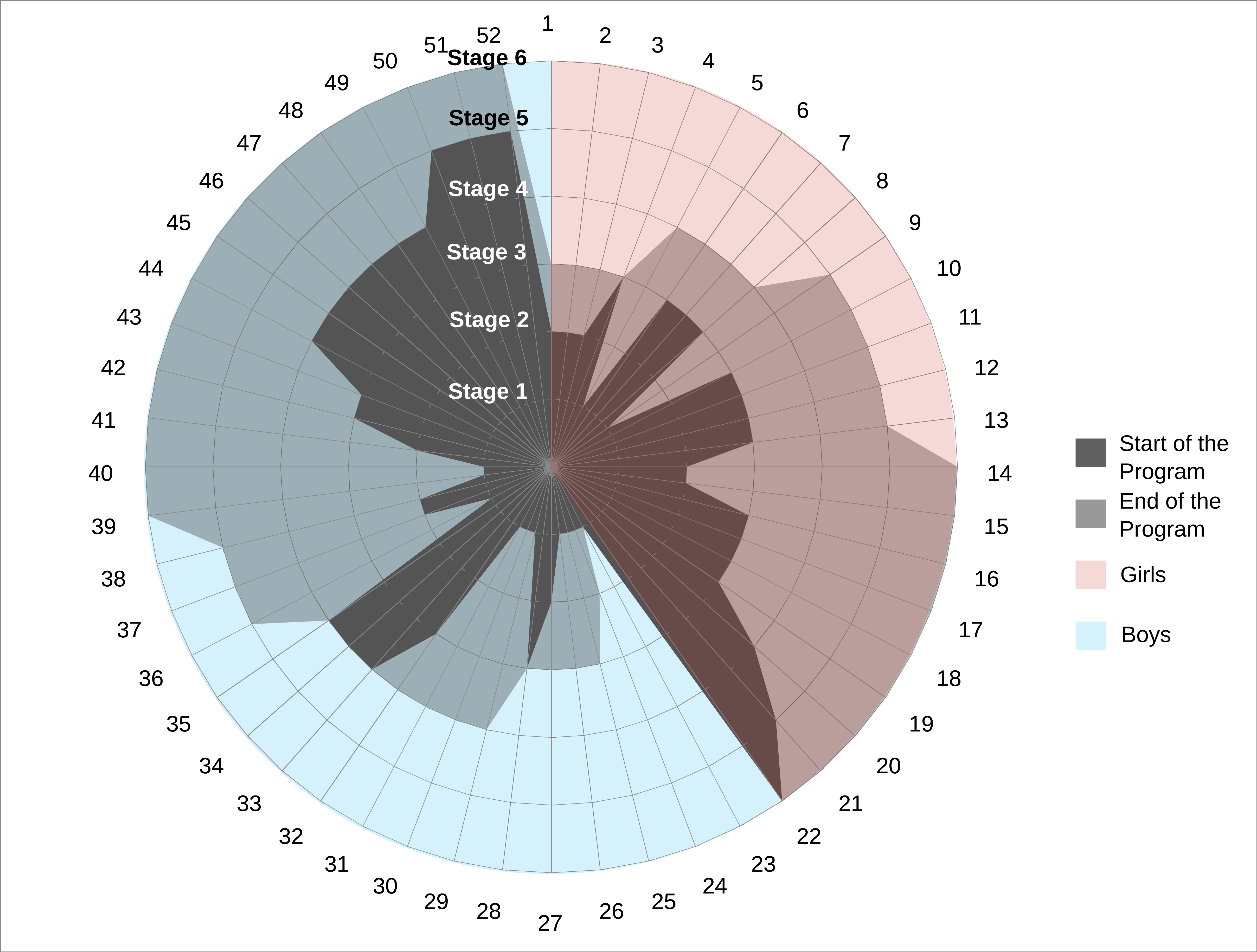}
\centering
\caption{Stages of usage of technology before and after the program}
\label{fig:stagesadoption}
\centering
\end{figure}

Fig.\ref{fig:stagesadoption} shows the radar plot with the children's responses on the Russell's stage that most appropriately described their stage of adoption of advanced technologies at the beginning and at the end of the program \cite{russell}. The plot, split between boys and girls, shows the increase in technology adoption before and after the program as perceived by the children. Seven students (13\%) did not think that their stage of technology usage had improved, all the other participants (83\%) reported one or more levels of improvement. We conducted a Friedman test to compare self reported technology adoption scores before and after the program. The results show a significant difference, ${\chi}^2(1) = 45$, p = 0.00 in technology adoption scores. Friedman test also showed significant different for boys ( ${\chi}^2(1) = 25$, p = 0.00), girls ( ${\chi}^2(1) = 20$, p = 0.00), rural ( ${\chi}^2(1) = 24$, p = 0.00), semi-urban ( ${\chi}^2(1) = 11$, p = 0.00) and urban ( ${\chi}^2(1) = 5$, p = 0.03).

\textbf{H3: Program would impact children's perceived competence, autonomy and relatedness equally regardless of gender and location}

The third finding from our STEAM engineering design thinking program is that the participants reported high competence, autonomy and relatedness from the program. Independent samples t-test showed that there was no statistically significant difference between perceived competence of girls (4.25 ± 0.44) and boys (4.17 ± 0.50), t(50)=0.62, p=0.54. There was also no statistically significant difference between rural (4.19 ± 0.52) and urban schools (4.20 ± 0.42), t(45)=0.05, p=0.96. Likewise, there was no statistically significant difference between perceived autonomy of girls (4.25 ± 0.78) and boys (4.22 ± 0.58), t(50)=0.18, p=0.86. There was also no statistically significant difference between rural (4.35 ± 0.59) and urban schools (4.03 ± 0.76), t(45)=1.60, p=0.12. 

Our analysis also showed that there was statistically significant difference between perceived relatedness of girls (4.82 ± 0.33) and boys (4.53 ± 0.51), t(50)=2.3, p=0.03. The statistical significance in perceived relatedness between girls and boys tells us that girls found the program to be more emotionally supportive and less stressful than the boys. We observed during the program that a few of the boy's teams had significant conflicts and disagreements among the team members which affected the overall team dynamics. However, the support extended by the teachers and mentors helped some of those teams to recover and complete their working prototypes successfully, while a few boys quit the program. This is reflected in the children's feedback. There was no statistically significant difference between the experiences of rural (4.58 ± 0.53) and urban school children (4.75 ± 0.32), t(45)=1.16, p=0.25. 

To sum it up, STEAM based Engineering Design Thinking programs on real world problem solving have a strong motivational aspect because they strongly influence the universal psychological needs of competence, autonomy and relatedness through a new social context, such as an extended after-school training program. Such programs communicate powerful lived experiences to adolescent children that shape their beliefs about themselves. To quote a girl participant \textit{``From dreaming to realization of it, ATL Marathon made the difference. This training program truly enlarged my vision of AI (Artificial Intelligence) and made me believe that nothing is impossible! Surely this program is recommended for each and every kid. I strongly believe that each and every student has a silent capability and talent hidden somewhere, which could be explored and expressed. Platforms of this magnitude serves well to at least cater a few. Hats off to the entire team who made it possible for me. Thank you very much from the bottom of my heart to make me believe that my dreams can be realized."} (Saavya)

\textbf{Challenges faced during program execution} Due to the Covid pandemic and the resulting school closures, we had to conduct nearly 80\% of the program online. This posed many challenges because several children had access only to smartphones and not to laptops or desktop computers at their home. To workaround this problem, mentors taught children alternative smartphone-based methods to use design and prototyping tools that were primarily designed for laptop or desktop usage. Despite being slower, many resourceful children could develop and demonstrate working digital prototypes by just using a smartphone. 

The online nature of the program made it difficult for children to collaborate with their team mates. We encouraged them to use online collaborative tools like Google Jamboard and Miro boards to make collaboration easier. Also, working online on team projects made it harder for the teams to debug their hardware issues. Towards the last three weeks of the program, we invited student teams to come in small batches to their ATL Makerspace following Covid-19 protocols and complete their projects.


\section{Conclusion}

Engineering design thinking is a creative process of identifying a problem, defining the problem, designing a solution, and creating innovation in the form of a working prototype. When integrated with social problem solving for UN SDGs, it gives a powerful learning experience for all children regardless of gender and school locale. Our results provide evidence that such programs offer equal learning opportunities for children in technical and non-technical skills, improve their perceived abilities in the use of advanced technologies such as artificial intelligence and IoT to design solutions for the greater good and bolster their psychological need for competence, autonomy and relatedness. Our research shows that even young pre-university students can think about complex systems and their interconnections and design innovative solutions. Such a constructionist approach to project-based learning builds student motivation and student engagement. It opens up avenues for children towards purposeful social interactions that reshape their self-beliefs. 

\section*{Acknowledgment}

We wish to express our gratitude to Sri Mata Amritanandamayi Devi, a world-renowned humanitarian and spiritual leader without whose guidance and undying support this research project would not have been possible. Many thanks to the mentor team comprising of researchers from Amrita Vishwa Vidyapeetham and alumni for helping us conduct this mentorship program. We would also like to thank the teachers, principals, and students of Amrita Vidyalayam schools for participating in this mentorship program and providing their valuable feedback.


\end{document}